\documentclass[runningheads]{llncs}
\usepackage{graphicx}
\usepackage{algorithm} 
\usepackage{algpseudo code} 
\usepackage{algorithmicx} 
\usepackage[dvipsnames]{xcolor}
\usepackage[T1]{fontenc}

\begin{document}
\title{A Rule-Based EEG Classification System for Discrimination of Hand Motor Attempts in Stroke Patients}
\author{Xiaotong Gu\and
Zehong Cao}
\authorrunning{X. Gu and Z. Cao}
\institute{University of Tasmania, Hobart TAS 7005, Australia\\ 
\email{\{xiaotong.gu,zehong.cao\}@utas.edu.au}\\}
\maketitle              
\begin{abstract}
Stroke patients have symptoms of cerebral functional disturbance that could aggressively impair patient's physical mobility, such as freezing of hand movements. Although rehabilitation training from external devices is beneficial for hand movement recovery, for initiating motor function restoration purposes, there are still valuable research merits for identifying the side of hands in motion. In this preliminary study, we used electroencephalogram (EEG) datasets from 8 stroke patients, with each subject involving 40 EEG trials of left motor attempts and 40 EEG trials of right motor attempts. Then, we proposed a rule-based EEG classification system for identifying the side in motion for stroke patients. In specific, we extracted 1-50 Hz power spectral features as input features of a series of well-known classification models. The predicted labels from these classification models were measured by four types of rules, which determined the finalised predicted label. Our experiment results showed that our proposed rule-based EEG classification system achieved $99.83 \pm 0.42 \% $ accuracy, $ 99.98 \pm 0.13\% $ precision, $ 99.66  \pm 0.84 \% $ recall, and $ 99.83 \pm  0.43\% $ f-score, which outperformed the performance of single well-known classification models. Our findings suggest that the superior performance of our proposed rule-based EEG classification system has the potential for hand rehabilitation in stroke patients.

\keywords{Stroke Rehabilitation \and Hand Motor Attempts \and EEG \and Classification.}

\end{abstract}
\section{Introduction}
Stroke is the second most common cause of death worldwide and the third most common cause of disability \cite{feigin2017global}. Clinical symptoms could be caused by acute signs that originate from global or focal brain dysfunction, possibly leading to severe impairment in patients’ mobility \cite{buch2008think,party2012national}. During the first year after a stroke occurs, a third of stroke patients have deficient or nonexistent remaining hand function, such as freezing of hand movement, and rare cases showed significant functional movement recovery in the following years \cite{buch2008think}. Unfortunately, since impairments are often resistant to therapeutic intervention, even after extensive conventional therapies, the probability of regaining functions for the impaired hand is low \cite{yue2017hand,kwakkel2003probability,lum2012robotic}.

A substantial amount of recent studies focusing on applying motor imagery (MI), the mental procedure of imagining body movements without physical actions \cite{scott2018motor}, to post-stroke hand function restoration, and review of MI study for post-stroke rehabilitation shows that MI may have value for patients recovering from stroke \cite{guerra2017motor}. There are also researches in which MI is used for stroke patients to control robotic exoskeleton for hand rehabilitation \cite{ferguson2018design,abdallah2017design}. MI-controlled robotic exoskeletons can be applied on the paretic hand to exercise for hand and upper limb, or as a 'third hand' to facilitate stroke patients to achieve daily functional practice \cite{ramadhan2019classification,yurkewich2019hand}. The MI-controlled robotic hand could be utilised for paretic patients who have impaired mobility of both hands, but there are differences between robotic exoskeleton and real hand, and for stroke patients, it is crucial to focus on restoring the functionality of the impaired hand. When utilising MI for hand rehabilitation for stroke patients compared with healthy subjects, cautions must be taken related to the side of the affected hand \cite{kemlin2016motor}, and it is possible to classify hand movement status in MI tasks \cite{ojeda2017classification}. Therefore, we believe applying hand-in-motion classification to stroke patients' hand movements, cooperating with MI technique for post-stroke hand restoration could have better outcomes for rehabilitation.

Compared with traditional clinical brain signal measurement technology, the recently fast-growing non-invasive brain-computer interface (BCI) brain signal monitoring devices are more accessible and effective for daily post-stroke rehabilitation in the home environment to practice for hand function restoration \cite{trujillo2017quantitative}. Electroencephalogram (EEG)-based BCI has been used in a combination of MI for stroke and neuro-rehabilitation \cite{ang2016eeg}. For example, the neuro-feedback mechanism, teaching self-control of brain functions to subjects by measuring brain waves and providing a feedback signal, can achieve a significant training performance for therapy and rehabilitation purposes. Here, we consider the EEG-based BCI \cite{gu2020eeg}, one of the non-invasive neuro-feedback mechanisms, to detect hand movement with EEG devices and assist stroke patients during the rehabilitation stage \cite{mohanty2018machine}. Currently, the EEG-based BCI studies have become a key approach for devising modern neuro-rehabilitation techniques \cite{chaudhary2016brain,chowdhury2018active,cao2018exploring,cao2018identifying}, and demonstrated promising feasibility for assisting in multiple healthcare domains \cite{cao2015classification,cao2019extraction,cao2019multi}. Abnormal EEG complexity of the brain, measured by entropy \cite{cao2017estimation,cao2017inherent,cao2019effects}, found in patients with acute stroke, showed an increased mean entropy value \cite{liu2016abnormal}. Current investigations focus on developing EEG-based BCI devices for hand function rehabilitation in stroke patients, and improving the performance of BCI classifiers to be applied for hand exoskeleton of stroke rehabilitation \cite{chowdhury2017online}. For instance, the eConHand showed 79.38\% classification accuracy for controlling the light hand exoskeleton in outpatient environments \cite{qin2019econhand}. Thus, we believe that the EEG-based BCI could be broadly used for hand movement detection and stroke rehabilitation clinically and in-home environment \cite{stroke2017robotic}. 

In this study, we explored a hand movement experiment and investigated the discrimination of hand motor attempts in stroke patients, by processing EEG signals collected from 8 stroke patients with impaired hand functions. In specific, the EEG data were firstly collected when the subjects were requested to conduct unilateral motor attempts, and then the EEG signals were labelled as "left" or "right" indicating the side of subjects' motor attempts. Then, the raw EEG data were transferred into power spectra, which were used for examining if exists significant differences of EEG power spectra between the left and right motor attempts and inputting EEG power features for training classifiers. The classifiers included in this study involved five well-known two-class classifiers and one rule-based classifier, possibly to improve the performance results and decrease the possibility of false predicted labels. The rule-based classifier is considered to be more interpretative and intuitive \cite{riid2017design}, and could improve the performance accuracy rate. 

The main contributions of our proposed rule-based EEG classification system comprise the following three parts: 
\begin{itemize}
\item The first study explored hand movement experiment of EEG signal for hand in motion attempts classification.
\item The proposed rule-based classification model achieved a comparatively high performance, which outperformed the performance of single well-known classification models.
\item The proposed post-stroke hand-in-motion classification system is beneficial to be used in combined with MI technique for stroke patients' hand rehabilitation in future research and practice.
\end{itemize}

\section{Materials and Method}

\subsection{Participant and EEG Data Recording}
\begin{figure}
\includegraphics[width=\textwidth]{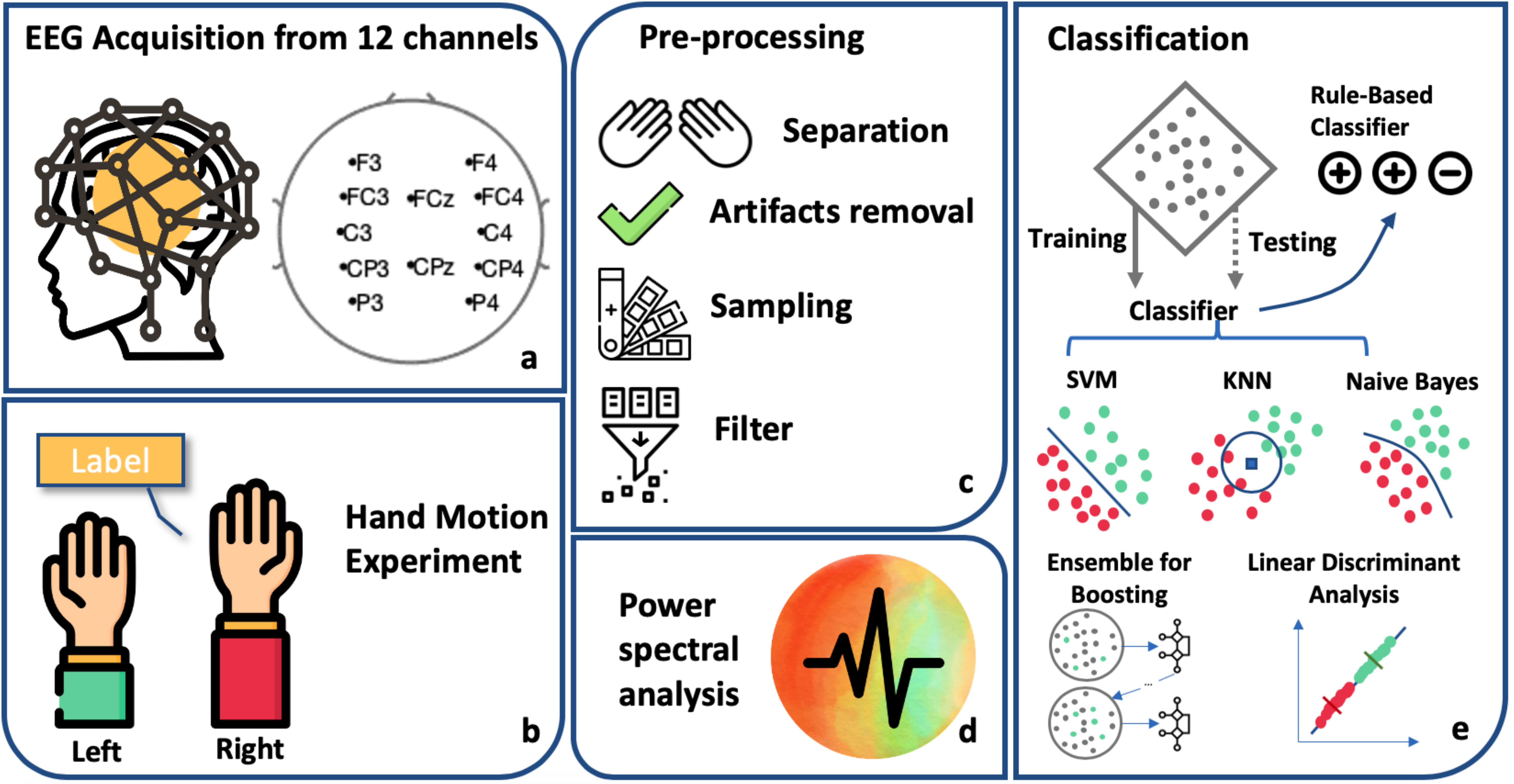}
\caption{Experiment and Data Analysis} \label{fig1}
\end{figure}

In this study, we used an EEG brain signal dataset \footnote{Link: https://github.com/5anirban9} of 8 hemiparetic stroke patients who have impaired functionality with either by left or right hand. Each subject conducted an equal number of left motor attempts as right motor attempts in the 80 trials with a sampling rate of 512 Hz. Each motor trail lasts 8 seconds, so the total numbers of each subject are 320 (8 seconds * 40 trails) for the left side and 320 (8 seconds * 40 trails) for the right side. Fig.~\ref{fig1} illustrates the procedure of the experiment, the analytical process of EEG data, and the applied classification models. The EEG signals of the motor attempts were recorded simultaneously from 12 channels (F3, FC3, C3, CP3, P3, FCz, CPz, F4, FC4, C4, CP4, and P4) according to the 10-20 international system as shown in Fig.~\ref{fig1}-a.

\subsection{EEG Data Analysis} 
All EEG data files were processed and analysed with EEGLAB in MATLAB software (The Mathworks, Inc.). EEGLAB is an extensible MATLAB toolbox for processing EEG and other electrophysiological signal data \footnote{https://sccn.ucsd.edu/eeglab/index.php} that offers interactive plotting functions, time/frequency transforms, artifact removal, and other extensions for EEG processing. 

\subsubsection{EEG data pre-processing}
The raw EEG data files we processed are firstly labelled for each trial as number "1" representing "right motor attempt" or number "2" representing "left motor attempt" as shown in Fig.~\ref{fig1}-b. Then, as shown in Fig.~\ref{fig1}-c, the raw EEG data were separated as left or right motion data based on the labels and filtered through 1Hz high-pass and 50 Hz low-pass finite impulse response filters as the recording sample rate is 512 Hz. The filtered data were then checked for any artifacts that need to be removed before being processed. Since no artifacts of visible muscle, eye-blink or other visible electromyography activity were detected, we proceeded processing on the filtered data. EEG signals have weak time-frequency-spatial characteristics, non-stationary, non-linear, and weak intensity, so to extract adaptive features reflecting frequency and spatial characteristics, it is critical to adopt feature extraction methods \cite{kim2018effective}. For this study, we converted the time-domain EEG data into the frequency domain and extracted power spectral features for left and right side motor attempts, as shown in Fig.~\ref{fig1}-d. We used the 256-point Fast Fourier Transforms (FFTs) window, which was set at 256-point data length, and in each window, the segment was converted into frequency domain respectively. 

\subsubsection{Statistical Analysis}
Before designing and applying classification models, since the EEG power spectra of each individual and cross-subjects are sufficient to conduct statistical analysis to determine if there is a significant difference between left and right motor attempts, we applied paired t-test to each frequency and channel of single-subject EEG power to determine the mean difference between the two sides. The \emph{p-value} of the paired t-test sets under 0.05, indicating the significant difference level of the left and right motor attempts of the stroke subjects. 

\section{Classification Models}
\subsection{Classification Models and Metrics}
Since EEG power has been labelled for each trial, we used supervised machine learning-based classification approaches, where one training sample has one class label \cite{zhou2017multi}, to train the motor attempt classifiers. With the cross-validation measurement, we set the three folds to randomly select two portions from each side motion features as the training set and the third portion from each side motion features as the testing set for the classifiers. The two training sets from left motions and the two training sets from the right motions were combined as the training sets for each classifier, while the testing set from left motions and the testing set from the right motions were combined as the testing set to be applied to each classifier. The training data labels were attached to the determining feature sets and then applied to the train classifiers, as shown in Fig.~\ref{fig1}-e. 

In this study, we used five well-known classification methods as classifiers which are Support Vector Machine (SVM), k-nearest neighbours (KNN), Naive Bayes, Ensembles for Boosting, and Discriminant Analysis Classifier \cite{mehmood2017optimal}. Each side's training sets of extracted power spectral features were provided to each classifier for training. The performance of each classifier was evaluated by applying the testing set to the trained classifiers to obtain the accuracy results. We also employed precision, recall, and F-score performance metrics to assess the performance of each classifier. 

\subsection{An EEG Rule-Based Classifier}
The rule-based classifier is considered to be more interpretative and intuitive than other existing classification methods \cite{riid2017design}. Because multiple classification models were involved in this study, to utilise fuzziness for mechanism and to acquire the maximum classification accuracy, we proposed a rule-based classification method to improve the classification model. As shown in Tab~\ref{tab1}, a single If-Then rule that if $x$ is $A$ then $y$ is $B$ is applied as an assumption in the form. The top three accurate classifiers are ranked as $ 1^{st} $, $ 2^{nd} $, and $ 3^{rd} $. If the classifier with the highest accuracy rate states positive, then the rule-based label is positive, while in all other cases, the rule-based label is negative. 

The pseudo-code of our proposed EEG rule-based classifier shows the discrimination of hand motor attempts in stroke patients as follows.

\begin{figure}
\includegraphics[width=\textwidth]{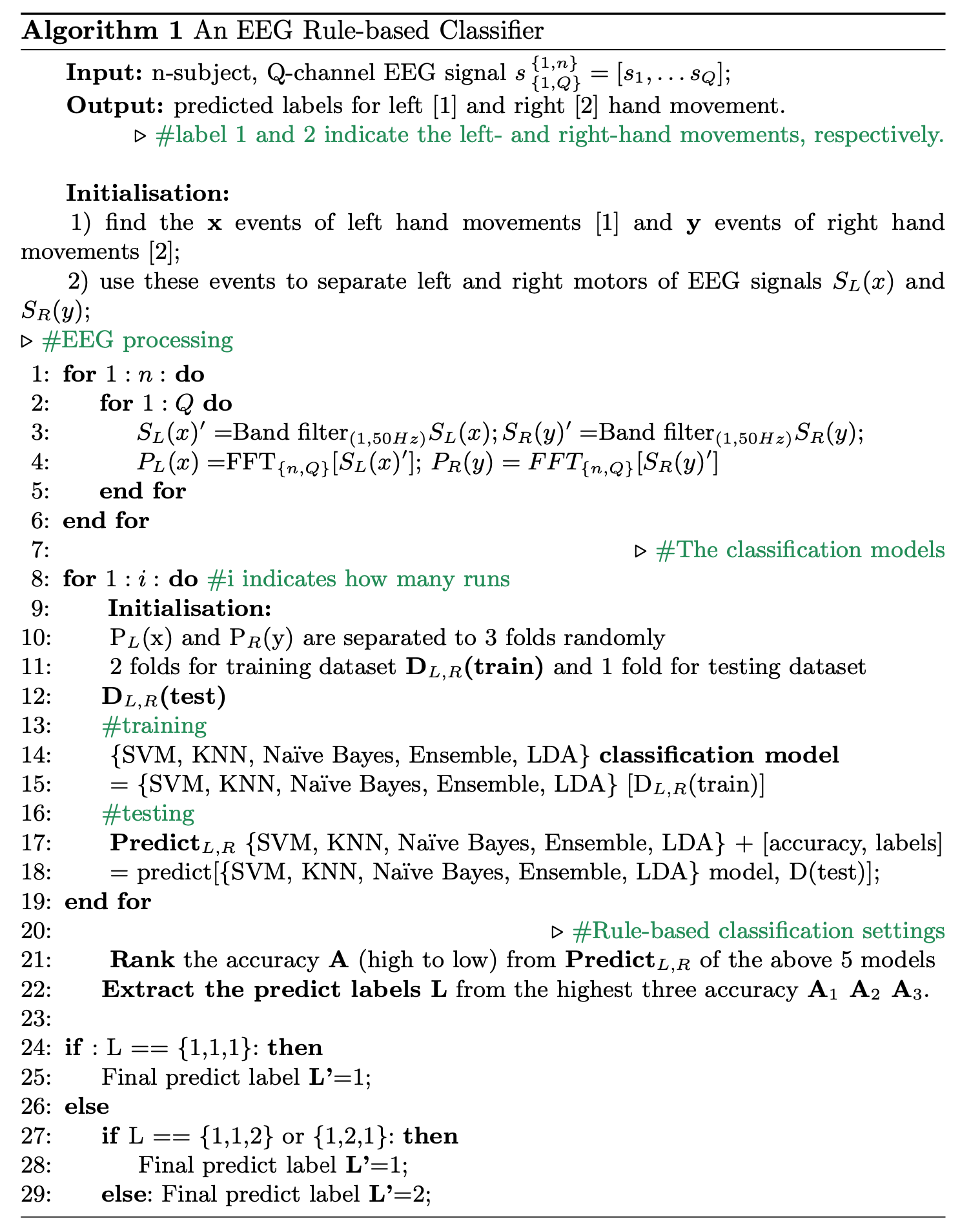}
\end{figure}

\begin{table}[htbp]
\caption{A Rule-based Classifier}
\begin{center}
\begin{tabular}{|c|c|c|c|c|}
\hline
\textbf{Predicted Labels} & \textbf{\textit{Classifier}}& \textbf{\textit{Classifier}}& \textbf{\textit{Classifier}}& \textbf{\textit{Rule-based Labels}} \\
\hline
Ranking-Accuracy& $ 1^{st} $& $ 2^{nd} $& $ 3^{rd} $&  \\
 & \multicolumn{3}{c}{IF}& THEN \\
Rule 1 &Positive &Positive &Positive & \\
Rule 2 &Positive &Positive &Negative &  Positive\\
Rule 3 &Positive &Negative &Positive & \\
Rule 4 &\multicolumn{3}{c}{Other Cases} &Negative \\ 
\hline
\end{tabular}
\label{tab1}
\end{center}
\end{table}

\section{Results}
For this study, we present two groups of findings which are the feature-based results of the EEG power spectra and the label classification performances of the five classifiers plus a rule-based classifier trained by the training sets and evaluated by the testing sets of 8 stroke patients. The results of this study are presented from two perspectives as power spectral feature-based and classifier-based.

\begin{figure}
\includegraphics[width=\textwidth]{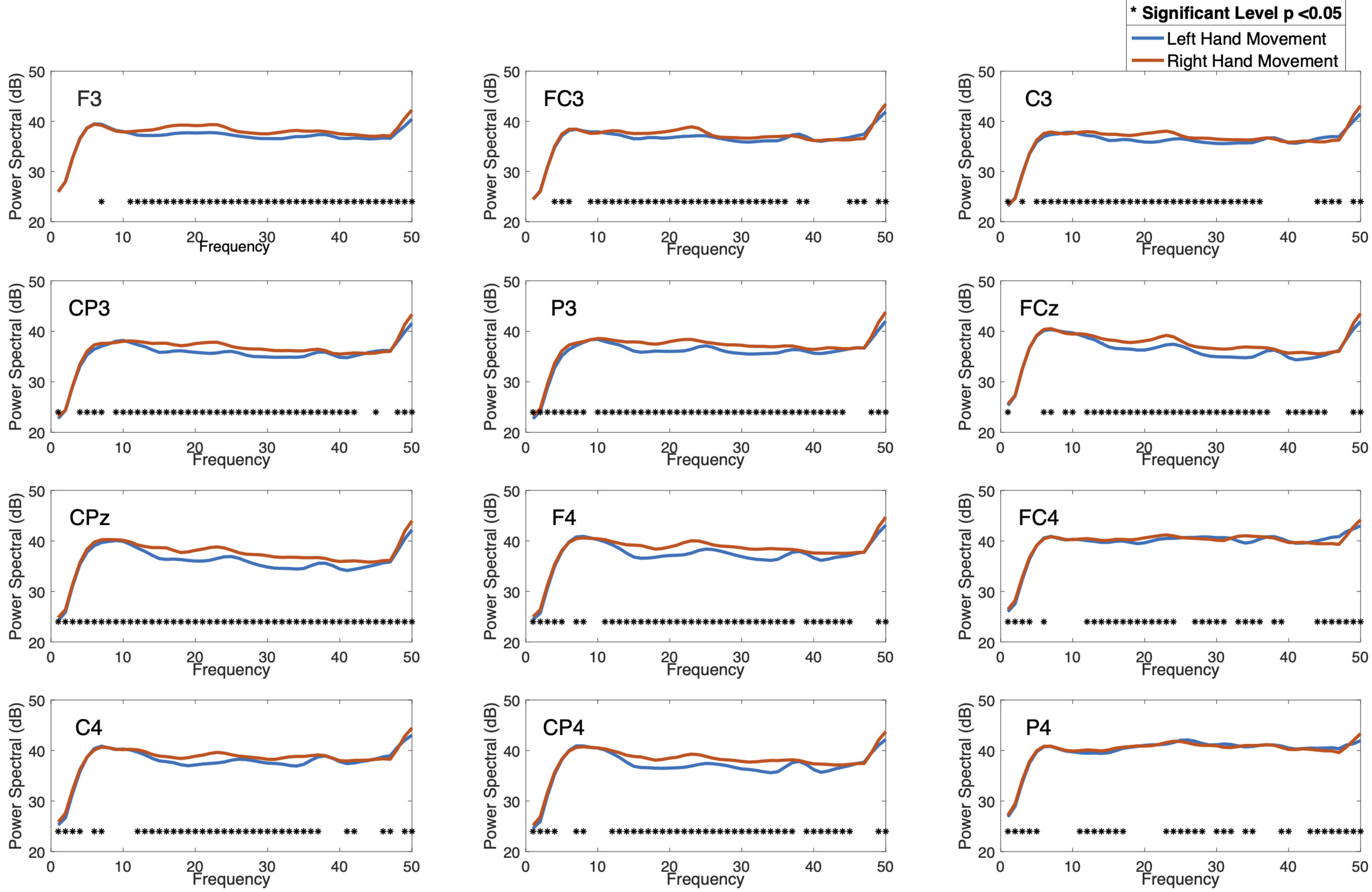}
\caption{Power Spectra between Movements of Left and Right Hands} \label{fig2}
\end{figure}

\begin{figure}
\includegraphics[width=\textwidth]{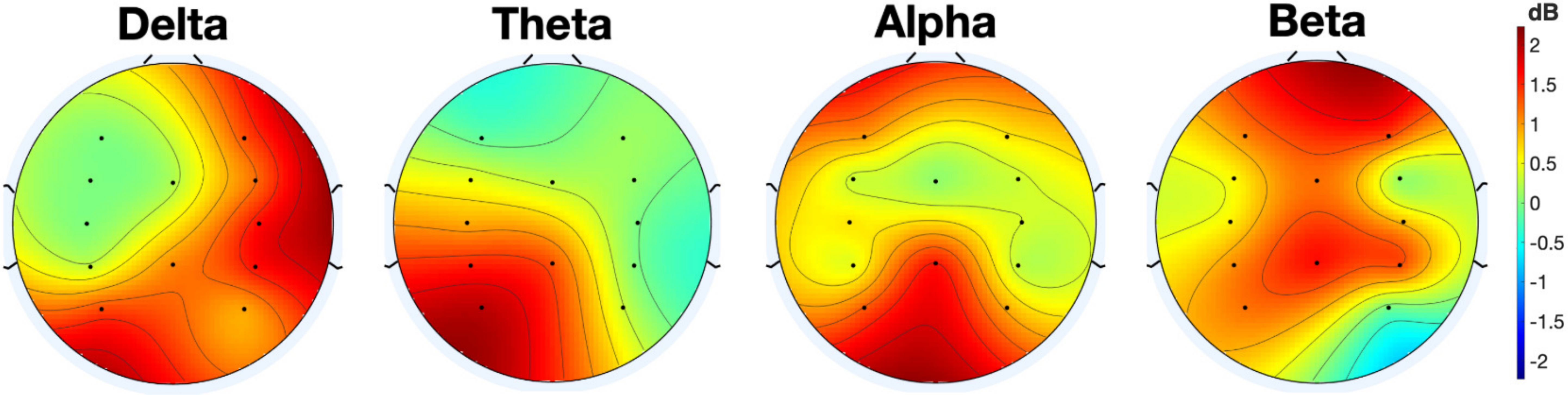}
\caption{Brain Scale Map Plot of Significant Difference of Power Spectra between Left and Right Motor Attempts} \label{fig3}
\end{figure}

We separated and plotted the power spectra of left and right motor attempts to inspect if there are significant differences of EEG signals for left and right-hand motions. As shown in Fig.~\ref{fig2}, the power spectra of left (blue colour) and right (red colour) motor attempts are demonstrated with the significant differences presented as black 'asterisk'. We also calculated the mean value of the significant difference of power spectra for each frequency and channel between left and right motor attempts in four frequency bands waves (delta, theta, alpha, and beta). For each of the 12 channels, the power spectra escalate rapidly and reach a peak at the frequency around 5 to 10 Hz, and maintained comparatively steady until surge to the highest power spectra level at 50 Hz. For 10 out of 12 channels (expect FC4 and P4), when the signal frequency is between approximately 10 to 40 Hz, the power spectra of left motor attempts and right motor attempts have the most significant differences ($p < 0.05$). These feature results are aligned with the tendency of power spectra mean value for the four frequency bands waves (delta, theta, alpha, and beta) of the 12 channels, as shown in Fig.~\ref{fig3}. For left and right motor attempts, on average, the most significant differences ($p < 0.05$) of power spectra mean value appear close to channel C4 in delta frequency range (3 Hz or lower), channel P3 in theta range (3.5 to 7.5 Hz), channel CPz in alpha range (7.5 to 13 Hz), and channel F4 in beta range (14 Hz to greater). The significant differences of the four frequency bands waves are demonstrated on brain scale map plotted with the 12 channel locations in Fig.~\ref{fig3}, and the calculation is as follow, where $ P_\Delta $ representing the significant difference (p $ < $ 0.05) of power spectra between left and right motion, movement of the left and right hands, $ P_R $ stands for power spectra of right motion, and $ P_L $ stands for power spectra of left motion. 

The classifier results are demonstrated in Tab~\ref{tab2} with the five well-known classifiers we used in this study for sub-results and the performance of a rule-based classifier we proposed as the improved results. Among the five trained existing classifiers, SVM performs the best in accuracy, precision, and F-score, while Linear Discriminant Analysis ranks the first for recall. Ensemble for Boosting classifier comes to third in accuracy, precision, and F-score, while KNN classifier ranks the third. Compared to the four top-ranking classifiers with performance above 98\% in all categories, Naive Bayes ranks the lowest at approximately 80\% for all four performance indicators. By employing a rule-based classifier we proposed in this study, we generally improved the proposed system's accuracy rate and the three performance metrics. Because the performance of most trained classifiers has an exceedingly high rate, for instance, SVM obtains $99.80 \pm 0.46 \% $ in accuracy and $99.97 \pm 0.16 \% $ in precision, the improvement rates of our proposed rule-based classifier are comparatively subtle. The experiment results show that our proposed rule-based EEG classification system can achieve $99.83 \pm 0.42 \% $ accuracy, $ 99.98 \pm 0.13\% $ precision, $ 99.66  \pm 0.84 \% $ recall, and $ 99.83 \pm  0.43\% $ f-score, and outperform the performance of single existing classification model.

{\centering
$ P_\Delta (p < 0.05) $ = $ P_R $ - $ P_L $\par
}

\begin{table}[htbp]
\caption{Classification Performance (left vs. right motor attempts)}
\begin{center}
\begin{tabular}{|c|c|c|c|c|}
\hline
\textbf{ Classifier(\%) }& \textbf{\textit{Accuracy}}& \textbf{\textit{Precision}}& \textbf{\textit{Recall}}& \textbf{\textit{F-score}} \\
\hline\hline
SVM &$99.80\pm 0.46$ &$99.97\pm 0.16$ &$99.64\pm 0.88$ &$99.81\pm 0.46$ \\
\hline
KNN &$98.33\pm 0.95$ &$98.41\pm 1.37$ &$98.28\pm 1.44$ &$98.33\pm 0.95$ \\
\hline
Naive Bayes &$79.52\pm 2.95$ &$78.55\pm 4.50$ &$80.23\pm 3.53$ &$79.30\pm 3.07$ \\
\hline
Ensemble for Boosting &$98.40\pm 1.06$ &$98.75\pm 1.28$ &$98.07\pm 1.51$ &$98.40\pm 1.05$ \\
\hline
Linear Discriminant Analysis&$99.73\pm 0.48$ &$99.75\pm 0.53$ &$99.72\pm 0.79$ &$99.73\pm 0.48$ \\
\hline
A Rule Classifier&$99.83\pm 0.42$ & $99.98\pm 0.13$ & $99.66\pm 0.84$ & $99.83\pm 0.43$ \\
\hline
\end{tabular}
\label{tab2}
\end{center}
\end{table}

\section{Discussion and Conclusion}
Stroke is one of the most common causes of disability, and stroke survivors commonly suffer impaired mobility. In recent years, an increasing number of researches applied EEG for MI-based BCI systems to stroke patients' hand rehabilitation, especially with the assistance of robotic or exoskeleton. There is also a difference between the affected and unaffected cerebral hemispheres correlated with the restoration of impaired hand's motor function. Therefore the significance of classifying which side of hands is in motion via EEG signal processing can not be neglected for stroke rehabilitation. In general, due to brain lateralisation, the discrimination between left and right hand MI is comparatively subtle, therefore high classification accuracy rate can be achieved for different sides of MI movement \cite{hashimoto2013eeg}. For stroke patients who suffer brain lesions, the significance of distinguishing affected and unaffected hand, and the difference between stroke patients and healthy people for interhemispheric interaction can not be neglected. The significance of our study supports the theory that stroke patients suffers from motor hemisphere dysfunctions, which reflects in EEG activities when conducting motor attempts. This finding is supportive to the conclusion in \cite{stkepien2011event} that affected hemisphere showed a smaller alpha event-related desynchronization compared to the unaffected hemisphere when each was contralateral to the acting hand. Another notable finding is that as shown in Fig.~\ref{fig3}, our study supports EEG power difference between left and right movement hand in stroke patients. This finding verifies the conclusion in \cite{saes2019does} that stroke patients present asymmetry in spectral power between hemispheres related to upper extremity motor function deficit. These notable findings suggests that EEG-based BCI systems could have significant contribution for hand motor rehabilitation in stroke patients.

In summary, in this study, we proposed a rule-based EEG classification system for categorising hand motor attempts in stroke patients. The predicted labels from these classification models were measured by four types of rules with a series of state-of-art classification models. The experiment results show that our proposed rule-based EEG classification system exceeds the performance of many single classification models for left and right motor attempts classification, with an accuracy rate of 99.83\%, precision rate of 99.98\%, recall rate at 99.66\%, and F-score rate at 99.83\%. The results of this proposed classification system indicate its feasibility in facilitating further EEG-based BCI systems for stroke rehabilitation, and the possibly promising outcome for cooperating EEG-based BCI hand-in-motion classification model with MI for post-stroke rehabilitation. 

\bibliographystyle{splncs04}
\bibliography{mybibliography}
\end{document}